\documentclass[sigconf]{acmart}
\usepackage{subcaption}
\makeatletter
\let\ACM@origcopyrightpermission\footnotetextcopyrightpermission
\renewcommand\footnotetextcopyrightpermission[1]{%
  \ACM@origcopyrightpermission{\setbox\z@\vbox{#1}%
    \vbox to \dimexpr\ht\z@+\dp\z@\relax{\vfil}}}
\def\copyrightpermissionfootnoterule{\kern-3\p@\kern2.6\p@}
\makeatother
\AtBeginDocument{%
\fancypagestyle{standardpagestyle}{%
  \fancyhf{}%
  \fancyhead[L]{\footnotesize PREPRINT}%
  \fancyhead[R]{\footnotesize\thepage}}
\fancypagestyle{firstpagestyle}{%
  \fancyhf{}%
  \fancyhead[L]{\footnotesize PREPRINT}%
  \fancyhead[R]{\footnotesize\thepage}%
  \fancyfoot[C]{\fbox{\parbox{0.55\textwidth}{\footnotesize This work has been
    submitted to the ACM/IEEE for possible publication. Copyright may be transferred
    without notice, after which this version may no longer be accessible.}}}}%
\pagestyle{standardpagestyle}}

\begin{document}

\title[RFChipAgent: Multi-Agentic AI Flow for Analog/RF Chip Design]{RFChipAgent: Multi-Agentic AI Flow for Analog/RF Chip Design}

\author{Awani Khodkumbhe}
\affiliation{%
  \institution{Qualcomm Technologies, Inc., \\ UC Berkeley}
  \city{}
  \state{}
  \country{}}
\email{}

\author{Yunfei Feng}
\affiliation{%
  \institution{Qualcomm Technologies, Inc.}
  \city{}
  \state{}
  \country{}}
\email{}

\author{Raj Rangarajan}
\affiliation{%
  \institution{Qualcomm Technologies, Inc.}
  \city{}
  \state{}
  \country{}}
\email{}

\author{Kevin Wang}
\affiliation{%
  \institution{Qualcomm Technologies, Inc.}
  \city{}
  \state{}
  \country{}}
\email{}

\author{Kamal Sahota}
\affiliation{%
  \institution{Qualcomm Technologies, Inc.}
  \city{}
  \state{}
  \country{}}
\email{}

\renewcommand{\shortauthors}{Khodkumbhe et al.}

\begin{abstract}
The wireless connectivity standards --- from Wi-Fi 7 to the emerging 6G frontier --- place stringent performance demands on Analog and RF circuits, which remain the critical interface between digital computation and the physical world. Yet, the Analog/RF chip design process remains one of the most labor-intensive and time-consuming. This work introduces a first-of-its-kind multi-agentic AI flow RFChipAgent for end-to-end design automation of Analog/RF circuits, in which autonomous AI agents collaboratively orchestrate the complete design flow under human supervision. RFChipAgent is built around four technical pillars. First, a multimodal retrieval-augmented generation (RAG) subsystem with private per-document FAISS indexing enables extraction of design knowledge from existing engineering documentation. Second, a topology agent drives topology selection, and a schematic and testbench agent automates circuit and testbench assembly. Third, a closed-loop hybrid sizing engine combines Tree-structured Parzen Estimator (TPE) and Covariance Matrix Adaptation Evolution Strategy (CMA-ES), evaluating every candidate in a simulator-in-the-loop framework. Fourth, a trust-scored simulation database accumulates verified performance data across trials and builds an adaptive optimization model that progressively informs subsequent trials. Together, these pillars replace the traditional trial-and-error design methodology with a systematic, self-improving agentic workflow. We validate RFChipAgent on a family of GF22FDSOI 60 GHz wideband Low Noise Amplifier (LNA) topologies, demonstrating automated topology generation, specification-driven design-space exploration, and simulator-guided optimization. Experimental results show substantial reductions in design effort while maintaining signoff-quality verification. This work establishes a foundation for next-generation multi-agentic AI flows for analog/RF electronic design automation.

\end{abstract}


 \maketitle

\section{Introduction}
The design of analog/RF integrated circuits remains a lengthy and laborious process because topology selection, device sizing, biasing, and passive-network design are tightly coupled. For example, in a 60 GHz Low Noise Amplifier (LNA), a change in transformer resonance or transistor finger count can substantially alter circuit performance. Recent AI-driven chip design approaches have shown promising directions toward automation, including topology generation \cite{eva}, multi-agent design frameworks \cite{multiagent}, reinforcement learning for optimization \cite{rose}, and LLM-based design reasoning \cite{reasoning}. However, an end-to-end multi-agentic analog/RF design AI flow remains unrealized due to the lack of a trustworthy framework that can seamlessly integrate design knowledge, topology synthesis, simulation-driven optimization, and continual learning.

\begin{figure*}[t]
  \centering
  \includegraphics[width=0.8\textwidth]{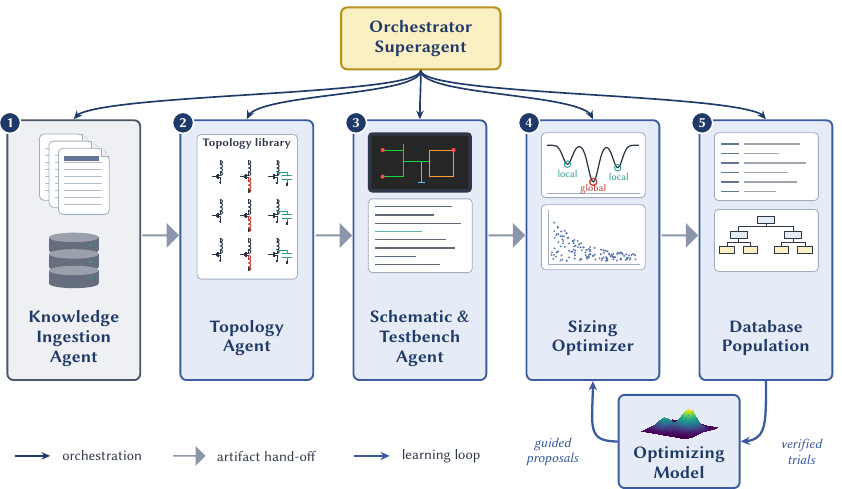}
  \caption{ Illustration of the proposed RFChipAgent flow.}
  \label{fig:flow}
\end{figure*}

As shown in Fig. \ref{fig:flow}, the proposed framework is orchestrated by a superagent that coordinates topology selection, automated schematic and testbench generation, simulator-in-the-loop optimization, and trust-scored knowledge accumulation. A retrieval-augmented generation (RAG) subsystem supports the topology agent by providing domain-specific design knowledge extracted from existing documentation. A key requirement for a trustworthy AI design flow is ensuring that invalid schematic assumptions, erroneous design modifications, and non-improving optimization actions are detected and filtered before they can propagate through subsequent design iterations.

The contributions are as follows:
\begin{itemize}
\item To the best of our knowledge, RFChipAgent is the first multi-agentic AI flow for end-to-end analog/RF circuit design automation, spanning knowledge ingestion, topology selection, schematic and testbench generation, simulator-in-the-loop optimization, and trust-scored knowledge accumulation.

\item A multimodal RAG subsystem with per-document FAISS indexing enables reliable retrieval of design knowledge while mitigating cross-document hallucinations.

\item A physics-aware topology agent that follows causally ordered design decisions, including stage selection, device selection, and matching-network synthesis, coupled with an automated schematic and testbench generation agent.

\item Physical-validity gates, including DC operating-point and stability checks, that are executed prior to objective evaluation, preventing invalid designs from propagating through the optimization loop.

\item A trust-scored simulation database that stores only verified design evaluations and uses accumulated performance data to progressively improve subsequent optimization iterations.
  \end{itemize}

\section{Preliminaries}

\subsection{Generative Analog/RF Design}

Recent work such as EVA \cite{eva} treats analog topology discovery as a generative task, using a transformer to predict device-pin connections and fine-tuning with Proximal Policy Optimization (PPO) or Direct Preference Optimization (DPO) for specification-driven topology generation. Multi-agent generative synthesis frameworks further separate topology generation from inverse design \cite{multiagent}. While these approaches demonstrate the promise of generative AI for topology exploration, they do not provide a complete design flow spanning knowledge ingestion, schematic generation, simulation-driven optimization, and trusted knowledge accumulation.
\subsection{Optimization and Reasoning}
RoSE \cite{rose} combines Bayesian optimization and reinforcement learning to improve the efficiency of analog parameter optimization under PVT variation. Reasoning-based LLM approaches for hardware generation have demonstrated that structured revision and compiler feedback can improve RTL correctness and reduce generation errors \cite{reasoning}. These works illustrate promising directions for introducing learning and reasoning capabilities into chip design workflows.
\section{Proposed RFChipAgent Flow}
In the proposed RFChipAgent flow, a superagent coordinates a knowledge-ingestion agent, a topology agent, a schematic and testbench generation agent, a sizing optimizer, and a trust-scored database. Accumulated simulation data guides optimization, while an LLM-in-the-loop provides recommendations for subsequent design campaigns, enabling a self-improving design flow.
\subsection{Knowledge Ingestion Agent}
The use of LLMs for design knowledge retrieval introduces the risk of hallucination, where a model may fabricate performance metrics, topology descriptions, or design insights that are not grounded in source documents. Such errors can silently propagate into downstream topology-selection decisions. Prior work such as MuaLLM \cite{muallm} mitigates this challenge by decoupling retrieval from inference through a hybrid RAG agent and ReAct workflow. However, these approaches typically rely on a single vector database spanning all ingested documents. While effective for broad retrieval, a unified index can introduce a more subtle provenance challenge in which performance data retrieved from one document may be incorrectly associated with a topology described in another.

\begin{figure}[t]
  \centering
  \includegraphics[width=0.8\linewidth]{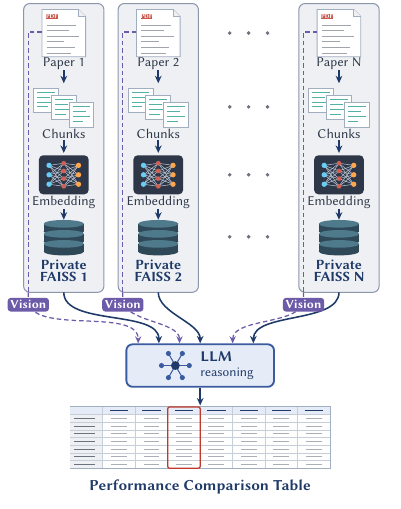}
  \caption{Proposed RAG framework for knowledge ingestion.}
    \label{fig:rag_new_diagram}
\end{figure}

To address this issue, the proposed RAG framework adopts a provenance-preserving retrieval strategy based on per-document indexing. As shown in Fig. \ref{fig:rag_new_diagram}, each PDF is independently parsed, chunked using a recursive text splitter, embedded using a private embedding model, and stored in its own FAISS index. A temporary combined search space is constructed only at query time. Because every retrieved chunk retains its source-document identity throughout retrieval and reasoning, topology descriptions, performance metrics, and design insights remain explicitly traceable to their original references. This design significantly reduces the risk of cross-document attribution errors that can arise in conventional merged-index RAG systems.

For the LNA case study, the retrieved design knowledge directly influenced the topology agent's architectural decisions. Specifically, the agent selected a transformer-based matching network for operation across the 60 GHz band, leveraging evidence from prior designs that demonstrated wideband impedance matching at mm-wave frequencies. In contrast, a conventional LC matching network would provide a narrower operating bandwidth and would be less suitable for the target specifications. The extracted knowledge base comprises ten LNA design references spanning a range of architectures. From these references, the most influential design priors were: (1) transformer-based matching is a widely adopted solution for wideband 60 GHz operation; (2) prior GF22FDSOI designs targeting the 60 GHz bands demonstrate the suitability of the technology node for mm-wave LNAs near the target specification range; and (3) when gain is the performance-limiting metric, the output stage often dominates overall gain and therefore warrants additional design emphasis. These retrieved priors provided an evidence-based foundation for topology selection and reduced reliance on unsupported design assumptions.

\subsection{Topology Agent}
The topology agent converts the target LNA specifications into a schematic design strategy using the retrieved design knowledge and the structured RF design rules. Although validated on LNAs in this work, the agent architecture is extensible to other analog/RF circuit classes through block-specific rules. Candidate topologies are filtered using physics-based validity checks prior to optimization. For the LNA case study, the following checks are applied.

\textbf{Stage Count}: For the LNA example, the agent uses PDK priors and power consumption budget to estimate whether a 2-stage design can meet the gain targets before adding a third stage. For the 15 mW power budget, a 2-stage design is preferred;
3-stage escalation is allowed only after the 2-stage loop exhausts its budget and the critic skill reports an unresolvable gain blocker.

\textbf{Active Selection}: In the LNA example, the first stage is selected primarily for noise. The agent encodes the characterization of the GF22FDSOI device and treats the cascode as an isolation/gain mechanism. It also captures the lessons from the existing database. 

\textbf{Passive Selection}: For the LNA case, the input network is assigned for NF and $S_{11}$; interstage and output transformers are assigned for $S_{21}$ and edge-band flatness. It also captures the lessons from the existing database.

\textbf{Prior Consistency}: Retrieved priors as well as database, if any, are used to initialize ranges. For example, in the LNA case study the knowledge-ingestion layer identifies transformer-based fourth-order networks as successful in wideband mm-wave LNAs, while the optimizer is still free to tune inductance and coupling.

These checks are applied in a sequential order: stage count is fixed first from the power budget, single-ended versus differential operation is decided next using range of power budget, each active device is then selected in turn (first stage for noise, later stages for gain), and only once every active device is final, the agent chooses the passive matching network.

\subsection{Schematic and Testbench Generation Agent} 
\begin{figure}[t]
  \centering
  \includegraphics[width=0.8\linewidth]{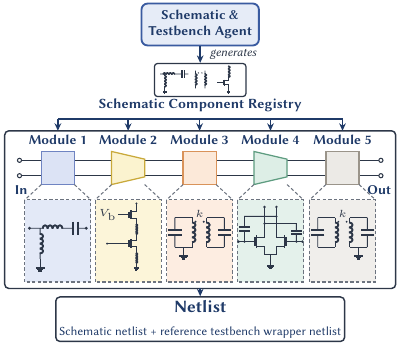}
  \caption{Schematic and testbench agent generating the design.}
    \label{fig:circuit}
\end{figure}

The RFChipAgent flow separates cell validation, schematic generation, and testbench assembly into independently auditable stages, as shown in Fig.~\ref{fig:circuit}. 

\textbf{Cell Validation}: Before schematic generation, a validation skill maintains inventories of active cells and passive matching-network cells available within the design library. Active cells include neutralized differential pairs, cascodes, and common-source stages, etc. while passive cells include transformers, T-networks, $\pi$-networks, and L-match structures, etc. All active cells expose a standardized 10-pin interface and all passive cells expose a standardized 7-pin interface, allowing cells to be interchangeably instantiated within a common module. Each candidate cell is verified for schematic existence, netlist availability, and schematic-validation status. Cells that fail any verification step are flagged and automatically rebuilt until a verified implementation is available, preventing invalid library elements from propagating into downstream design stages. 

\textbf{Schematic Generation}: Once the topology agent produces a sequence of active and passive modules, the schematic-generation agent automatically instantiates and connects the selected cells to form the target LNA architecture. A topology-consistency check is subsequently performed to verify connectivity and architectural correctness. Detected violations trigger automatic reconstruction attempts before the design is admitted to the optimization flow. 

\textbf{Testbench Assembly}: The testbench-generation agent constructs a simulation-ready environment using a reference testbench template containing simulator directives, port definitions, bias sources, and analysis configurations. The generated circuit netlist is automatically inserted into this template to create a complete verification setup. Missing or invalid upstream artifacts result in explicit preflight failures, ensuring that incomplete designs are detected before simulation and optimization.

\subsection{Sizing Optimizer}
The generated topology contains FETs that must be biased and sized, as well as matching networks whose component values must be optimized to satisfy the design targets. A multi-stage optimization flow is therefore employed.

\textbf{Sigmoid-Normalized Multi-Objective Function}:
Combining noise figure (NF), gain ($S_{21}$), input return loss ($S_{11}$), gain ripple, and DC power consumption ($P_{DC}$) into a single objective function is challenging for two reasons. First, these metrics exist on different physical scales (dB versus mW). Second, some specifications are maximization objectives ($S_{21}$) while others are minimization objectives (NF, $S_{11}$, and $P_{DC}$). Directly averaging raw values would distort the optimization process because numerically larger quantities could dominate the objective despite having similar margins relative to their specifications. Likewise, commonly used figures of merit (FoMs) do not fully capture circuit quality; for example, a passive wire could achieve a deceptively high FoM compared to an actual amplifier. Following the sigmoid-plus-harmonic-mean formulation of \cite{chenna}, each metric is first mapped independently onto a common quality scale between 0 and 1 using a sigmoid function,
\begin{equation}
\sigma(z,k)=\frac{1}{1+e^{-z/k}}
\end{equation}
where $z$ represents the margin relative to the specification and $k$ controls the transition steepness. The metric quality is defined according to the specification type: \begin{equation} Q_{\max}(x)=\sigma\!\left(x-x_{\mathrm{spec}},k\right), \qquad Q_{\min}(x)=\sigma\!\left(x_{\mathrm{spec}}-x,k\right). \end{equation} This formulation ensures that $Q\rightarrow 1$ when the design satisfies its specification and $Q\rightarrow 0$ when it violates the specification, regardless of whether the metric is a maximize-type or minimize-type objective. Accordingly, $Q_{S21}$ uses $Q_{\max}$, while $Q_{NF}$, $Q_{S11}$, and $Q_{PDC}$ use $Q_{\min}$. The flatness metric $Q_{flat}$ additionally penalizes in-band gain ripple once $S_{21}$ approaches its target value, ensuring ripple optimization is only emphasized when meaningful. The steepness parameter $k$ is selected individually for each specification according to its design criticality. Smaller values of $k$ produce a sharper transition centered at the specification boundary, creating a stronger local gradient for hard constraints. Larger values spread the gradient across a broader margin range, allowing the optimizer to differentiate varying levels of violation without immediate saturation. In the LNA validation example, $k_{NF}=0.4$ is used because NF permits little margin, whereas $k_{S21}=k_{S11}=2.0$ is chosen to preserve useful optimization signal over several decibels of gain or return-loss margin. The five quality metrics are then combined using a harmonic mean, \begin{equation} Q_{\mathrm{total}} = \frac{5} {\frac{1}{Q_{NF}} +\frac{1}{Q_{S21}} +\frac{1}{Q_{S11}} +\frac{1}{Q_{flat}} +\frac{1}{Q_{PDC}}}, \end{equation} rather than an arithmetic mean. Because the harmonic mean is dominated by its smallest term, it prevents designs from achieving a high overall score by excelling in only a subset of metrics while failing another. 

\textbf{Exploration and Exploitation Algorithms}: The circuit design space may contain up to 21 optimization variables (15 for the LNA in Section~\ref{sec:exp}) spanning transistor dimensions, bias conditions, and matching-network parameters. At this dimensionality, a single optimization strategy is inefficient. Therefore, a two-stage optimization approach is adopted. Tree-structured Parzen Estimator (TPE) optimization is first used for global exploration because it efficiently handles mixed continuous and discrete variables, including integer-valued transistor finger counts. Once a promising region has been identified, Covariance Matrix Adaptation Evolution Strategy (CMA-ES) is applied for local exploitation and fine-grained refinement of the continuous matching-network parameters around the best solutions discovered by TPE. 

\textbf{Per-Trial Physical Validity Gates}: An optimizer operating solely on a scalar objective cannot distinguish between a physically valid low-performing design and a physically invalid design that violates fundamental circuit requirements. As a result, optimization resources may be wasted exploring infeasible regions of the design space. To address this issue, two physical-validity checks are performed after each simulation trial and before objective-function evaluation. First, a DC operating-point check verifies the region of operation of every transistor. Any trial in which at least one device falls outside saturation or subthreshold operation is immediately discarded. Second, a stability check computes the Edwards--Sinsky $\mu$-factor from the extracted S-parameters over the frequency range from 1~Hz to 90~GHz. Trials are rejected if $\mu_{\min}<1$ or if the global $S_{11}$ exceeds 0~dB at any frequency within this range, indicating potential instability. These validity gates prevent physically infeasible solutions from consuming optimization budget and ensure that only realizable circuit candidates contribute to the search process. 

\subsection{Trust-Score Gated Database} RFChipAgent flow maintains a persisted database of every trial's parameter vector and outcome, partitioned into a \emph{good} set $\mathcal{G}$ and a
\emph{bad} set $\mathcal{B}$ by graded cost and trust status. Two non-parametric density estimates are fit over the free-variable space, $a(x)$ over $\mathcal{G}$ and $b(x)$ over $\mathcal{B}$, and a candidate point is preferred when the likelihood ratio $a(x)/b(x)$ is large: this identifies regions of the design space that have historically produced good outcomes relative to bad ones, rather than regions that are simply under-sampled. 

This model can be audited and reused across campaigns. However, the model is only as reliable as the labels feeding it. Therefore, a trial is only admitted to $\mathcal{G}$ or $\mathcal{B}$ if it clears the trust-score threshold. Every run is assigned a trust score with 70\% weight on execution integrity and
30\% on data provenance:
\begin{equation}
T(d)=0.7T_{exec}(d)+0.3T_{prov}(d).
\end{equation}
Execution integrity is computed as a weighted sum of verification indicators,
\begin{equation}
T_{exec}=\sum_i w_i I_i,
\end{equation}
where $I_i=1$ only when the corresponding check is satisfied. The checks are PSF availability, nonzero $P_{DC}$, full-band stability extraction, and optional corner/EM coverage. Provenance
separately scores whether each reported number comes from verified Spectre data, PDK-derived priors, estimates, or missing data.

Between campaigns, the LLM computes deterministic diagnostics and what-if analysis over the completed trial database---search stagnation since the last cost improvement, which free variables sit pinned at a search-space boundary in the top trials, and whether the best trial's margin on any unmet specification sits on the steep or flat region of that specification's sigmoid penalty. For boundary-pinned variables, the LLM helps in distinguishing a physical edge (propose widening the bound) from a nonphysical edge, such as a capacitance being pushed toward a negative value (propose removing that component from the topology instead, a judgment a numeric optimizer cannot make on its own). 

\begin{table*}[t]
  \caption{RFChipAgent flow top-5 proposed design candidates performance for GF22FDSOI 60 GHz wideband LNA.
  Specifications: \mbox{S\textsubscript{21} $\geq$ 25\,dB}, \mbox{NF $\leq$ 4.5\,dB},
  \mbox{S\textsubscript{11} $\leq$ $-$10\,dB}, \mbox{Ripple $\leq$ 3.5\,dB}, \mbox{P\textsubscript{DC} $\leq$ 15\,mW}}
  \label{tab:opt_results_new}
  \setlength{\tabcolsep}{5pt}
  \begin{tabular}{lcccccc}
    \toprule
    & \multicolumn{5}{c}{\textbf{Performance metrics}} & \\
    \cmidrule(lr){2-6}
    \textbf{Trial} &
      \textbf{S\textsubscript{21}} & \textbf{NF} & \textbf{S\textsubscript{11}} &
      \textbf{Ripple} & \textbf{P\textsubscript{DC}} & \textbf{Specification pass?} \\
    & (dB) & (dB) & (dB) & (dB) & (mW) & \\
    \midrule
    \textit{Specification} &
      $\geq 25$ & $\leq 4.5$ & $\leq -10$ & $\leq 3.5$ & $\leq 15$ & --- \\
    \midrule
    \,\#43  & 21.7 & 4.2 & $-$7.9  & 2.6 & 13.3 &
      No \\
    \,\#70  & 22.0 & 4.2 & $-$10.1 & 2.6 & 13.4 &
     No \\
    \,\#86  & 24.1 & 4.3 & $-$11.8 & 1.4 & 13.2 &
      No \\
    \,\#104 & 23.4 & 4.3 & $-$10.8 & 3.7 & 11.8 &
      No \\
    \textbf{\,\#174} & \textbf{27.1} & \textbf{4.2 }& \textbf{$-$16.9} & \textbf{3.5 }& \textbf{14.3 }&
      \textbf{Yes} \\
    \bottomrule
    \multicolumn{7}{l}{}
  \end{tabular}
\end{table*}

\begin{table}[t]
  \caption{RFChipAgent flow proposed design candidate variables for the GF22FDSOI
  60\,GHz wideband LNA (best trial\,\#174).}
  \label{tab:variables_cma}
  \setlength{\tabcolsep}{4pt}
  \begin{tabular}{lcc}
    \toprule
    \textbf{Description} & \textbf{Unit} & \textbf{Value} \\
    \midrule
    \multicolumn{3}{l}{\textit{Active device sizing}} \\
    \midrule
    Stage-1 FET finger count           & ---    & 28    \\
    Stage-2 FET finger count           & ---    & 16    \\
    Stage-1 gate inductance        & pH     & 174   \\
    Stage-1 noise-figure inductor  & pH     & 53.9  \\
    Stage-1 source degeneration inductor    & pH     & 50.1  \\
    Stage-1 gate-source coupling   & ---    & 0.094 \\
    \midrule
    \multicolumn{3}{l}{\textit{Bias currents}} \\
    \midrule
    Stage-1 bias current        & $\mu$A & 866   \\
    Stage-2 bias current           & $\mu$A & 295   \\
    \midrule
    \multicolumn{3}{l}{\textit{Interstage transformer}} \\
    \midrule
    Coupling factor     & ---    & 0.345  \\
    Primary inductor  & pH     & 285    \\
    Secondary inductor & pH     & 105    \\
    \midrule
    \multicolumn{3}{l}{\textit{Input auto-transformer}} \\
    \midrule
     Shunt inductor         & pH     & 211   \\
    \midrule
    \multicolumn{3}{l}{\textit{Output transformer}} \\
    \midrule
    Coupling factor         & ---    & 0.570   \\
    Primary inductor      & pH     & 220     \\
    Secondary inductor    & pH     & 169     \\
    \bottomrule
  \end{tabular}
\end{table}

\section{Experiments and Evaluations}\label{sec:exp}
\subsection{Experimental Setup}

The RFChipAgent flow is validated end-to-end on a GF22FDSOI two-stage LNA targeting
60\,GHz wideband operation, chosen because it exercises every stage of the
loop---topology selection, schematic\slash testbench assembly, closed-loop sizing optimizer, and trust-gated database---under the challenging specifications characteristic of RFICs for emerging wireless standards. The specification is $S_{21}$\,min $\geq 25$\,dB, NF\,max
$\leq 4.5$\,dB, $S_{11}$\,max $\leq -10$\,dB, in-band ripple $\leq 3.5$\,dB, and
$P_{DC}$ $\leq 15$\,mW. The topology agent selects a neutralized-differential-pair
second stage cascaded with a common-source degenerated first stage, matched with an input auto-transformer and two transformer-based interstage/output networks, consistent with the transformer-preference prior extracted by the knowledge-ingestion agent.
The sizing optimizer runs its TPE global phase followed by CMA-ES local polish with simulator-in-the-loop for every
trial; every completed trial is passed through the two per-trial validity gates to ensure correct DC operating point and stability. The trust-score is computed before
the data is written to the trust-gated database, so every number reported below is drawn only from trials that cleared both the physical-validity gates and the trust threshold, not from raw simulator output.

\enlargethispage{2\baselineskip}
\subsection{Performance Results}
\begin{figure}[t]
  \centering
  \includegraphics[width=\linewidth]{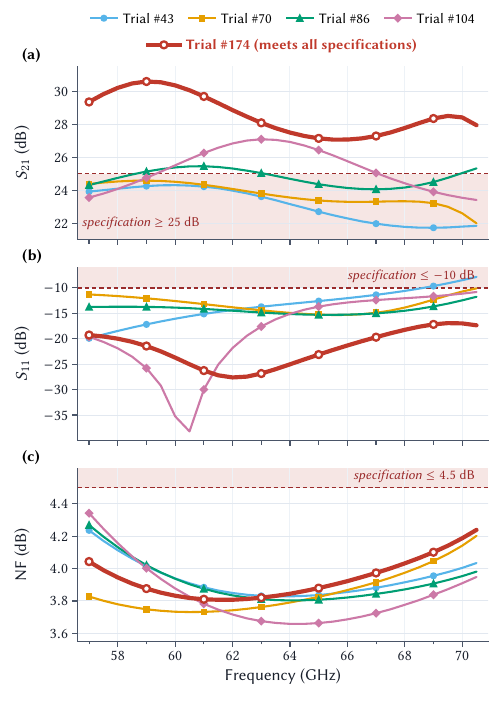}
  \caption{RFChipAgent flow output: performance of top 5 candidates.}
    \label{fig:perf5}
\end{figure}

Table~\ref{tab:opt_results_new} reports the five highest-ranked trials retained by the trust-gated database, alongside the specification for direct comparison. Trial\,\#174 meets every
specification simultaneously. Fig.~\ref{fig:perf5} shows the top five candidates' $S_{21}$, $S_{11}$, and NF traces across the full 60\,GHz band. Table~\ref{tab:variables_cma} lists every free variable at the Trial\,\#174 design point. 

\subsection{Efficiency and Computational Cost}
Table~\ref{tab:token_usage} reports the aggregate token usage and wall-clock time for RFChipAgent, attributed to superagent orchestrating every step in Fig.~\ref{fig:flow}---knowledge ingestion,
topology selection, schematic\slash testbench assembly, and
optimizer---end to end. The campaign completed in approximately 2 hours of wall-clock
time at an estimated cost of \$156.20. Compared to manual labor-intensive circuit design, this is a substantial improvement. 

\begin{table}[t]
\caption{RFChipAgent flow estimated cost and computation time}
\label{tab:token_usage}
\begin{tabular}{lr}
\toprule
\textbf{Metric} & \textbf{Value} \\
\midrule
Input tokens        & 10,756      \\
Cache write tokens   & 24,543,985  \\
Cache read tokens    & 152,703,166 \\
Output tokens        & 1,196,007   \\
Est. cost            & \$156.20    \\
Wall-clock time      & $\sim$2 hr  \\
\bottomrule
\end{tabular}
\end{table}

\section{Conclusion}
We presented a multi-agentic AI flow for end-to-end design automation of Analog/RF circuits. The flow decides a topology, generates schematic and assembles testbench as well as optimizes the circuit. It verifies every trial, and records results in a database. A private FAISS-based RAG subsystem supports the knowledge ingestion. Using a GF22FDSOI 60 GHz wideband LNA as validation example, the flow is able to meet target performance provided to it. RFChipAgent achieves this result in wall-clock time of around ~2 hrs with computation cost of \$156.20---a considerable improvement over manual design. RFChipAgent provides a practical path toward AI-native analog/RFIC design automation.

\begin{acks}
The authors thank Qualcomm Technologies, Inc. WRD WAN RFIC for discussions and Hasnain Lakdawala, Lincoln Leung, and Jeremy Dunworth for useful feedback.
\end{acks}


\begin{thebibliography}{10}

\bibitem{eva}
J. Gao, W. Fu, X. Guo, W. Cao, and X. Zhang. 2025. EVA: An Efficient and
Versatile Generative Engine for Targeted Discovery of Novel Analog Circuits.
ACM/IEEE Design Automation Conference (DAC).

\bibitem{multiagent}
S. Wang, Q. Li, H. He, J. Gao, Z. Wang, Y. Sun, X. Zhang, T. Chi, and W. Cao.
2025. Invited Paper: Multi-Agent Generative Synthesis for Analog/RF Circuit:
from Scalable Topology Generation to Efficient Inverse Design. IEEE/ACM
International Conference on Computer-Aided Design (ICCAD).

\bibitem{rose}
J. Gao, W. Cao, and X. Zhang. 2023. RoSE: Robust Analog Circuit Parameter
Optimization with Sampling-Efficient Reinforcement Learning. ACM/IEEE Design
Automation Conference (DAC).

\bibitem{reasoning}
W. Fu, S. Li, K. Yang, X. Zhang, Y. Jin, and X. Guo. 2025. Building Reasoning
LLMs for Hardware Design Generation via Function-Aligned Differentiated
Revision. IEEE/ACM International Conference on Computer-Aided Design (ICCAD).

\bibitem{chenna}
V. Chenna and H. Hashemi. 2026. A Nonintuitively Frequency-Staggered Wideband
mm-Wave Low-Noise Amplifier. IEEE International Solid-State Circuits
Conference (ISSCC).

\bibitem{muallm}
P. Abbineni, S. Aldowaish, C. Liechty, S. Noorzad, A. Ghazizadeh Ghalati, and
M. Fayazi. 2026. MuaLLM: A Multimodal Large Language Model Agent for Circuit
Design Assistance with Hybrid Contextual Retrieval-Augmented Generation. Asia South Pacific Design Automation Conference (ASP-DAC).

\end{thebibliography}
\end{document}